\documentclass[aps,pre,superscriptaddress,showpacs]{revtex4}

\usepackage[dvips]{epsfig}
\usepackage[dvips]{graphicx}
\usepackage{latexsym,amsmath,verbatim}
\usepackage[dvips,usenames]{color}

\newcommand{\av}[1]{\langle {#1} \rangle}

\begin{document}

\title{Glass transition and random walks on complex energy landscapes}

\author{Andrea Baronchelli} 
\affiliation{Departament de F\'\i sica i Enginyeria Nuclear, Universitat
  Polit\`ecnica de Catalunya, Campus Nord B4, 08034 Barcelona, Spain}

\author{Alain Barrat}
\affiliation{Centre de Physique Th\'eorique (CNRS UMR 6207), Luminy,
13288 Marseille Cedex 9, France}
\affiliation{Complex Networks Lagrange Laboratory,
Institute for Scientific Interchange (ISI), Torino, Italy}

\author{Romualdo Pastor-Satorras}
\affiliation{Departament de F\'\i sica i Enginyeria Nuclear, Universitat
  Polit\`ecnica de Catalunya, Campus Nord B4, 08034 Barcelona, Spain}
\date{\today}

\begin{abstract}
We present a simple mathematical model of glassy dynamics seen as a
random walk in a directed, weighted network of minima taken as a
representation of the energy landscape. Our approach gives a broader
perspective to previous studies focusing on particular examples of
energy landscapes obtained by sampling energy minima and saddles of
small systems. We point out how the relation between the
energies of the minima and their number of neighbors should be studied
in connection with the network's global topology, and show how the tools
developed in complex network theory can be put to use in this context.
\end{abstract}


%64.70.Q- 	%Theory and modeling of the glass transition 
%05.40.Fb 	%Random walks and Levy flights 
%64.60.aq 	%Networks
%89.75.-k 	%Complex systems
%89.75.Hc 	%Networks and genealogical trees 
\maketitle

The physics of glassy systems, the glass transition, and the slow
dynamics ensuing at low temperatures have been the subject of a large
interest in the past decades \cite{Debenedetti}. In particular,
special attention has been devoted to the dynamics of a glassy system
inside its configuration space: The idea is to understand glassy
dynamics in terms of the exploration of a complex, rugged energy
landscape in which the large number of metastable states limits the
ability of the system to
%find its ground state and 
equilibrate. In the picture of an energy landscape partitioned into
basins of attraction of local minima (``traps''), the dynamics of the
system is separated into harmonic vibrations inside traps and jumps
between minima \cite{Angelani:1998}. Several models of dynamical
evolution through jumps between traps have been proposed and studied
in order to reproduce the phenomenology of the glass transition,
pointing out various ingredients of the ensuing slow dynamics
\cite{Bouchaud:1992,Monthus:1996}. Moreover, several works have mapped
the energy landscape of small systems and studied the dynamics through
a master equation for the time evolution of the probability to be in
each minimum. Systems thus considered range from clusters of
Lennard-Jones atoms to proteins or heteropolymers
\cite{Angelani:1998,Cieplak:1998,Bongini:2008b,Carmi:2009}.

The success of these approaches has recently brought about a number of
studies focusing on the topology of the network defined by considering
the minima as nodes and the possibility of a jump between two minima
as a (weighted, directed) link \cite{newman-review}. The small-world
character of these networks has been pointed out \cite{Scala:2001}, as
well as strong heterogeneity in the number of links of each node
(its degree).  Scale-free distributions have been observed
\cite{Doye:2002}, and linked to scale-free distributions of the areas
of the basins of attraction \cite{Massen:2005,Seyed:2008}.
%, as well as to preferential attachment mechanisms \cite{Massen:2005}. 
Further investigations of various energy landscapes (of Lennard-Jones
atoms, proteins, spin glasses) have used complex network analysis
tools
\cite{Doye:2005,Bongini:2008b,Seyed:2008,Carmi:2009,network_as_a_tool}.
For instance, some works have exposed a logarithmic dependence of the
energy of a minimum on its degree
\cite{Doye:2002,Seyed:2008,Carmi:2009}, or energy barriers increasing
as a (small) power of the degree of a node \cite{Carmi:2009}. However,
the relation between the energy and the degree of a minimum has never
been systematically investigated. Moreover, no systematic study of the
connection between the network of minima and the glassy dynamics has
been performed, since the studies cited above are limited to 
small size systems.

Here, we make an important first step to fill this gap by putting
forward a simple mathematical model of a network of minima, through a
generalization of Bouchaud's trap model
\cite{Bouchaud:1992,Monthus:1996}. This framework allows to use the
wide body of knowledge developed recently on dynamical phenomena in
complex networks \cite{Barrat:2008} to study the dynamics in a complex
energy landscape as a random walk in a directed, weighted complex
network. The corresponding heterogeneous mean-field (HMF) theory
\cite{DorogoRev}
highlights the connection between network properties and  dynamics,
and shows in particular that the relationship between energy and
degree of the minima is a crucial ingredient for the existence of a
transition and the subsequent glassy phenomenology. This approach
sheds light on the 
fact that 
scale-free structures and logarithmic
relations between degrees and energies 
have been 
empirically found,
and should stimulate more systematic investigations on this issue. It
also puts previous studies of the dynamics in a network of minima
obtained empirically in a broader perspective.

We consider the well-known traps model of phase space consisting in
$N$ traps, $i=1,\cdots,N$, of random depths $E_i$ extracted from a
distribution $\rho(E)$ \cite{Bouchaud:1992,Monthus:1996}. The dynamics
is given by random jumps between traps: The system, at temperature
$T=1/\beta$, remains in a trap for a time $\tau_0 \exp(\beta E)$
(where $\tau_0$ is a microscopic timescale that we can set equal to
$1$), and then jumps to a new, randomly chosen trap; all traps are
connected to each other, in a fully connected topology.  Here we
consider instead a---more realistic---case in which the traps form a
network: Each trap $i$ has depth $E_i$ and number of neighbors $k_i$.
The system, pictured as a random walker in this network, escapes from
a trap of depth $E_i$ towards one of the $k_i$ neighboring traps of
depth $E_j$ with a rate $r_{i\to j}$, which is a priori a function of
both $E_i$ and $E_j$.  Possible rates include Metropolis
$\frac{1}{k_i}\min \left(1, e^{\beta(E_j-E_i)} \right)$ or Glauber
$k_i^{-1}/\left(1+ e^{-\beta(E_j-E_i)}\right)$ ones. For simplicity,
we will stick here to the original definition of rates depending only
on the initial trap, i.e. $r_{i\to j}=e^{-\beta E_i}/k_i$.

In the fully connected trap model, all traps are equiprobable after a
jump, so that the probability for the system to be in a trap of depth
$E$ is simply $\rho(E)$, and the average time spent in a trap is
$\langle \tau \rangle= \int \rho(E) e^{\beta E} dE$. Thus, a
transition occurs between a high temperature phase in which $\langle
\tau \rangle$ is finite and a low temperature phase with diverging
$\langle \tau \rangle$ if and only if $\rho(E)$ is of the form $\sim
\exp(-\beta_0 E)$ at large $E$ (else the transition temperature is
either $0$ or $\infty$) \cite{Monthus:1996}; the distribution of
trapping times is then $P(\tau)\sim \tau^{-1-T/T_0}$. Let us see how
this translates when the network of minima is not fully connected. In
this case, it is convenient to divide the nodes in degree classes, as
it is usual in the framework of the HMF theory \cite{DorogoRev}.  We
further assume that the depth of a minimum and its degree are related:
$E_i = f(k_i)$ where the function $f(k)$ does not depend on $i$ and is
a characteristic of the model. The time spent in a trap of degree $k$
is then $\tau_k = e^{\beta f(k)}$, and the transition rate $r_{i\to
  j}$ between two traps can be written as a function of the endpoints'
degrees $k_i$ and $k_j$. It is important to recall that, in the steady
state, the probability for a random walker to find itself on a node of
degree $k$ is $k P(k)/\langle k \rangle$, where $P(k)$ is the degree
distribution of the network and $\langle k \rangle$ is the average
degree \cite{Redner:2001}. The average rest time
%after 
before a hop is therefore
\begin{equation}
%\langle \tau \rangle = \sum_k \frac{kP(k)}{\langle k \rangle} e^{\beta f(k)} \ .
\langle \tau \rangle = \langle k \rangle^{-1}\sum_k kP(k) e^{\beta f(k)} \ .
\label{eq:tau}
\end{equation}
It is then clear that the presence of a finite transition
temperature at which $\langle \tau \rangle$ becomes infinite
results from an interplay between the topology of the underlying network 
%as given by $P(k)$, 
and the relation between traps' depth and
degree. For instance, for a scale free distribution $P(k) \sim
k^{-\gamma}$, a finite transition temperature is obtained \textit{if and only
if} $f(k)$ is of the form $E_0 \log(k)$: $\langle \tau \rangle$ is then
finite (in an infinite system) for $T > T_c \equiv E_0/(\gamma -2)$,
and infinite for $T \leq T_c$. For $P(k)$ behaving instead as
$e^{-ak^\alpha}$, $f(k)$ has to be of the form $E_0 k^\alpha$ for a
transition to occur. 
Thus, although important, the study of the topology of the network of
minima is not enough to understand the dynamical properties of the
system, and more attention should be paid to the energy/connectivity
relation.

To gain further insight into the dynamics of the system we can write,
within the HMF approach, the rate equation for the probability
$\rho_k(t)$ that a given vertex of degree $k$ hosts the random walker at
physical time $t$. Since the walker escapes a trap with rate per
unit time $r_k=1/\tau_k$, we have
\begin{equation}
  \frac{\partial \rho_k(t)}{\partial t} = -r_k \rho_k(t) + k \sum_{k'}
  P(k'|k) r(k'\to k) \rho_{k'}(t)
\label{eq:RE}
\end{equation}
where $P(k'|k)$ is the conditional probability that a random neighbor
of a node of degree $k$ has degree $k'$ \cite{newman-review}.  In the
steady state, $\partial_t \rho_k(t)=0$, the solution of Eq.~(\ref{eq:RE}) for
any correlation pattern $P(k'|k)$ is \cite{Redner:2001}
\begin{equation}
r_k \rho_k \sim  k \ ,
\end{equation}
and the normalized equilibrium distribution reads
\begin{equation}
  \rho_k^{eq} = \frac{k\tau_k}{N\langle k\tau_k \rangle}.
\label{eq:eq}
\end{equation}
Note that the probability for the random walker to be in {\em any}
vertex of degree $k$ is then $P_{eq}(k)=N P(k) \rho_k^{eq}$.  Since $\langle
k\tau_k \rangle=\sum_k kP(k) e^{\beta f(k)}$, the conclusion is the same as before: A
normalizable equilibrium distribution exists indeed if and only if $\langle
k \tau_k \rangle < \infty$, and the presence of a transition at a finite temperature
$T_c$ is determined by the interplay between $P(k)$ and $f(k)$.

% Approach to equilibrium

\begin{figure}[t]
%\begin{center}
\includegraphics*[width=0.42\textwidth]{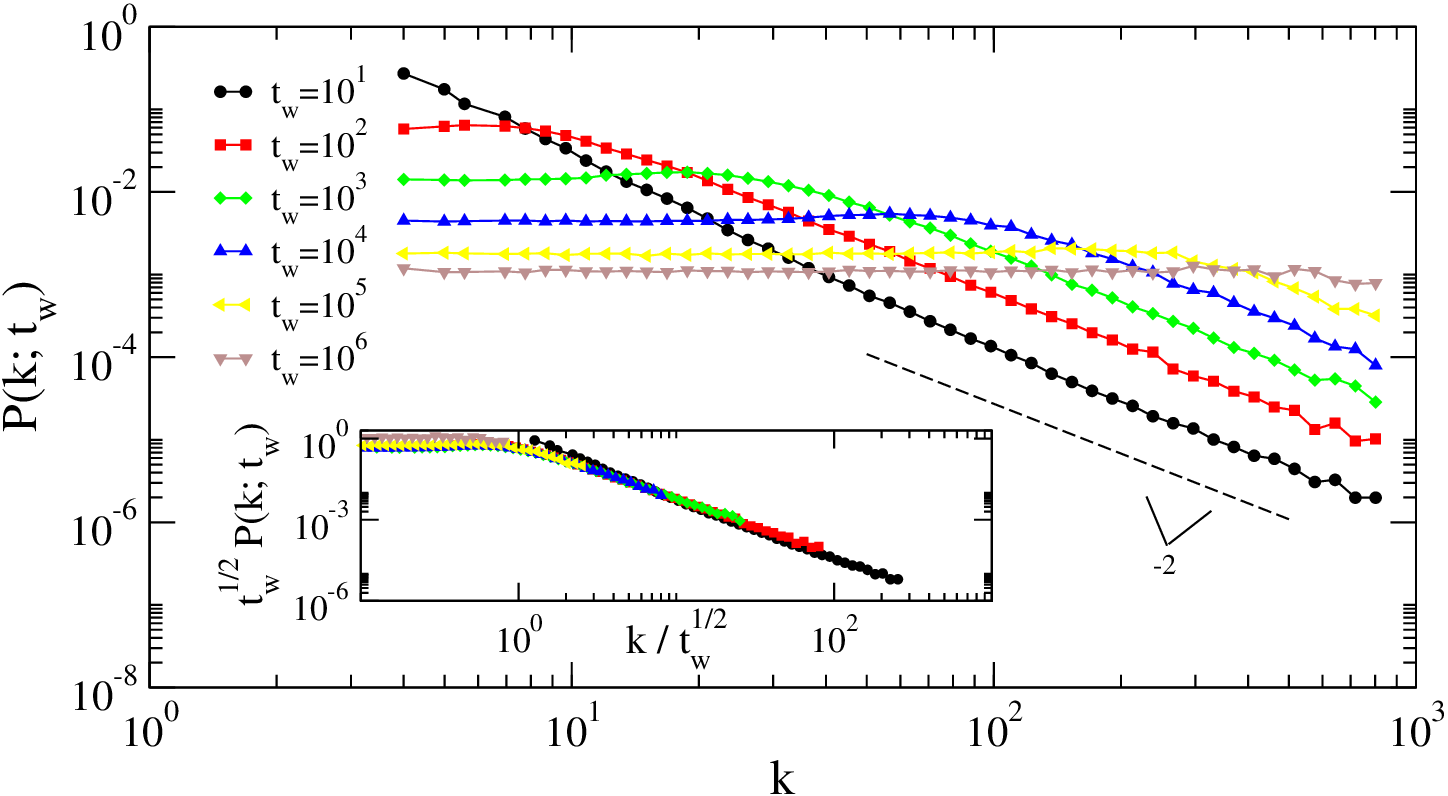}
%\end{center}
\caption{Evolution of $P(k;t_w)$ for an uncorrelated scale-free
  network.  Here $N=10^6$ ($k_c=10^3$), $\gamma=3$, $\beta E_0=2$ so that
  $P(k;t_w) \sim k^{-2}$ at short times and $P_{eq}(k) \sim k^{0}$. Inset:
  $t_w^{1/2}P(k;t_w)$ vs $k/t_w^{1/2}$ for $t_w < t_{eq} \sim
  10^6$.}
\label{fig:pk_tw}
\end{figure}

In any finite system, the distribution $P_{eq}(k)$ exists, and the
probability that the random walker is in a node of degree $k$ at time
$t_w$, $P(k;t_w)=N P(k) \rho_k(t_w)$, converges to $P_{eq}(k)$ after a
certain equilibration time. It is interesting to study this evolution
in the low temperature regime, when it exists. Let us consider the
case of a scale-free network with $P(k) \sim k^{-\gamma}$ \cite{newman-review}
and $f(k)=E_0 \log(k)$, i.e., $\tau_k =k^{\beta E_0}$. In numerical
experiments, the walker explores an underlying network generated
according to the Uncorrelated Configuration Model
\cite{Catanzaro:2005}, and spends in each node of degree $k$ an amount
of time extracted from the distribution $P(\tau_k) = \exp(-t/\tau_k)/\tau_k$.
Figure \ref{fig:pk_tw} shows how $P(k;t_w)$ evolves from the
distribution $kP(k)/\av{k}$ at short times, corresponding to the
degree distribution of a node reached after a random jump, to
$P_{eq}(k) \sim k^{1+\beta E_0-\gamma}$ (obtained from Eq.~(\ref{eq:eq})) at long
times: The small degree region equilibrates first, and a progressive
equilibration of larger and larger degree regions takes place at
larger times. Small degrees correspond in fact to shallow minima,
which take less time to explore, while large degree nodes are deep
traps which take longer to equilibrate~\footnote{The equilibration
  proceeds therefore in an ``inverse cascade'' from the small nodes to
  the hubs, while usual diffusion processes on networks (random walks,
  epidemics) visit first large degree nodes and then cascade towards
  small nodes \cite{Barrat:2008}.}. At time $t_w$, one can therefore
consider that the nodes of degree smaller than a certain $k_w$ are
``at equilibrium'', while the larger nodes are not. Considering that
the total time $t_w$ is the sum of the trapping times of the visited
nodes, which is dominated by the longest one $k_w^{\beta E_0}$, we obtain
$k_w \sim t_w^{1/(\beta E_0)}$ . Figure~\ref{fig:pk_tw} shows indeed that the
whole non-equilibrium distribution can be cast into the scaling form
\footnote{For $P(k) \sim e^{-k/m}$ and $f(k)=E_0 k/m$, the same reasoning
  applies with $k_w \sim \ln (t_w)$, and $P(k;t_w) \sim k e^{-k/m}$ for $k \gg
  k_w$, $P(k;t_w) \sim k e^{(\beta E_0-1)k/m}$ for $k \ll k_w$ (not shown).}
\begin{equation}
P(k;t_w)= t_w^{-1/(\beta E_0)} F\left(k / t_w^{1/(\beta E_0)} \right)
\label{eq:pktw_scaling}
\end{equation}
where $F$ is a scaling function such that $F(x)\sim x^{1+\beta
  E_0-\gamma}$ at small $x$ and $F(x)\sim x^{1-\gamma}$ at large $x$.
This evolution takes place until the largest nodes, of degree $k_c$,
equilibrate. For an uncorrelated scale-free network, $k_c \sim
N^{1/2}$ so that the equilibration time is $t_{eq} \sim k_c^{\beta E_0} \sim
N^{\beta E_0/2}$.

\begin{figure}[t]
%\begin{center}
\includegraphics[width=0.42\textwidth]{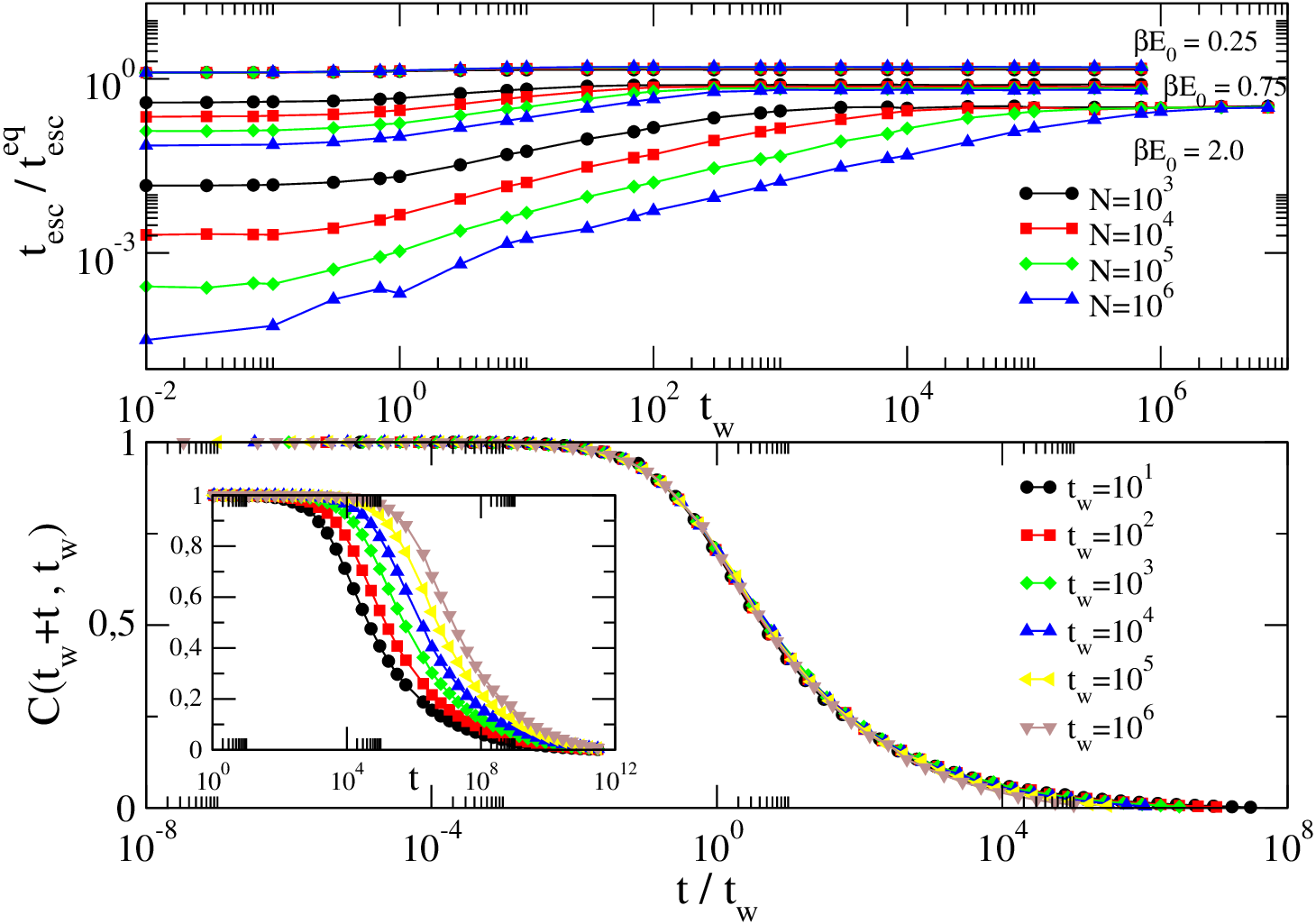}
%\end{center}
\caption{Top: average escape time $t_{esc}(t_w)$ divided by the large
  time prediction Eq.~(\protect\ref{eq:t_esc_m1}), for various $N$ and $\beta$.
  Bottom: $C(t_w+t,t_w)$ vs $t/t_w$ for an uncorrelated scale-free
  network of $N=10^6$ minima.  Here $\gamma=3$, $\beta E_0=4$. Inset:
  $C$ vs $t$.  }
\label{fig:tesc_C}
\end{figure}

The evolution of $P(k;t_w)$ at low temperature corresponds to the
aging dynamics of the system, which is exploring deeper and deeper
traps. This dynamics is also customarily investigated through a
two-time correlation function $C(t_w+t, t_w)$ between the states of
the system at times $t_w$ and $t_w+t$, defined as the 
the average probability that a particle has not changed trap between
$t_w$ and $t_w+t$~\cite{Bouchaud:1992,Monthus:1996}: this amounts to considering that the correlation is
$1$ within one trap and $0$ between distinct traps.
The probability that a walker remains in trap $i$
a time larger than $t$ is simply given by $\exp(-t/\tau_i)$, so that
\begin{equation}
C(t_w+t, t_w) = \int dk \; P(k;t_w) e^{-t/\tau_k}, 
\end{equation}
where we have used the continuous degree approximation,
replacing discrete sums over $k$ by integrals. For scale-free
networks, using the scaling form (\ref{eq:pktw_scaling}), it is then
straightforward to obtain that the correlation function obeys
the so-called ``simple'' aging $C(t_w+t, t_w)=g(t/t_w)$, as in the
original trap model \cite{Monthus:1996} (Fig. \ref{fig:tesc_C}).

Aging properties of the system can be measured also through the average
time $t_{esc}(t_w)$ required by the random walker to escape from the
node it occupies at time $t_w$. 
In other words we define $t_{esc}=\av{t'} -t_w$, where
$t'>t_w$ is the time of the first jump performed by the walker after
$t_w$, which gives $t_{esc}(t_w) = \int dk \; \tau_k P(k;t_w)$.  For
small $t_w$ with respect to the equilibration time, $t_{esc}$ is
growing due to the evolution of $P(k;t_w)$. At long enough times, in
any finite system, $\rho_k(t_w) \to \rho_k^{eq}$ so that
\begin{equation}
  t_{esc}(t_w \to \infty)=  \int dk \; \frac{k  P(k) e^{2\beta f(k)}}{\langle k\tau_k \rangle}.
\end{equation}
Interestingly, this formula shows that, whenever $P(k)$ and
$f(k)$ are such that a finite transition temperature $T_c$ exists,
$t_{esc}(t_w \to \infty)$ actually diverges at $2T_c$. The existence
of a diverging timescale at $2T_c$ was in fact  already noted in the
original mean-field trap model \cite{Monthus:1996}.

We can also consider how $t_{esc}$ diverges with the system size depending
on temperature. For instance, with $P(k) \sim k^{-\gamma}$ and $f(k)=E_0 \log(k)$, we have
$t_{esc}(t_w \to \infty) \equiv t_{esc}^{eq}= \langle k^{1+2\beta
  E_0}\rangle/\langle k^{1+\beta E_0}\rangle$. For uncorrelated
networks, the cut-off of $P(k)$ scales as $k_c \sim N^{1/2}$, so that
\begin{equation}
  t_{esc}^{eq} \simeq \left\{
    \begin{array}{lcl}
    N^{\beta E_0/2} & \mathrm{if} & \beta E_0>\gamma-2 \\
    \frac{N^{\beta E_0/2}}{\ln N} & \mathrm{if} & \beta E_0=\gamma-2 \vspace{0.1cm}\\
    N^{(2+2\beta E_0 -\gamma)/2} & \mathrm{if} & \frac{\gamma-2}{2}<\beta E_0<\gamma-2\\
    \ln N  & \mathrm{if} & \beta E_0 = \frac{\gamma-2}{2} \\
    \mathrm{const.} & \mathrm{if} & \beta E_0 < \frac{\gamma-2}{2}
  \end{array}
  \right. .
\label{eq:t_esc_m1}
\end{equation}
Figure~\ref{fig:tesc_C} displays a numerical check of these predictions.
For an exponential degree distribution $P(k) \sim e^{-k/m}$, 
with $f(k)=E_0 k/m$, we obtain analogously
$t_{esc}^{eq}= \av{k e^{\frac{2 \beta E_0 k}{m}}}/ \av{k e^{\frac{\beta E_0 k}{m}}}$, and,
considering that $k_c \sim m\ln N$ we obtain
\begin{equation}
  t_{esc}^{eq} \simeq \left\{
    \begin{array}{lcl}
    N^{ \beta E_0} & \mathrm{if} & \beta E_0 > 1 \\
    \frac{N^{2  \beta E_0 -1 }}{\ln N} & \mathrm{if} & \beta E_0=1 \vspace{0.1cm}\\
    N^{2  \beta E_0-1 } & \mathrm{if} & 1/2 <\beta E_0< 1\\
    \ln N  & \mathrm{if} & \beta E_0= 1/2 \\
    \mathrm{const.} & \mathrm{if} &  \beta E_0< 1/2
  \end{array}
  \right. .
\label{eq:t_esc_m1_caseB}
\end{equation}

We finally turn to the investigation of a quantity of particular
relevance in random walks on networks, namely the mean first passage
time (MFPT) \cite{Redner:2001}. 
Since the way in
which the phase space is explored is crucial for the dynamical
properties of the system, it is also interesting in the present
context to measure the MPFT averaged over random origin-destination
pairs, $\av{t_{MFPT}}$. This procedure was for instance used in \cite{Carmi:2009} to
extract a global relaxation time, whose temperature dependence was
tentatively fitted to a Vogel-Tammann-Fulcher form $\exp(A/(T-T_0))$,
with however $T_0 \ll T_c$. The framework put forward above allows in
fact to rationalize this result. The average number of hops performed
by a random walker between two nodes, $H_{MFPT}$, does not indeed depend on the
temperature. On the other hand, the temperature determines the
interplay between the physical time and the number of hops: the
time needed to perform $H$ hops is
\begin{equation}
\sum_{i=1}^H \tau_i, 
\end{equation}
where $\tau_i=e^{\beta f(k_i)}$ is the residence time in trap $i$. Therefore,
\begin{equation}
\av{t_{MFPT}} = H_{MFPT} \langle \tau \rangle,
\end{equation}
where $\langle \tau \rangle$ depends on temperature, $P(k)$ and $f(k)$
as given by Eq.~(\ref{eq:tau}). Let us consider the concrete example of
an uncorrelated scale-free network with degree distribution $P(k) \sim
k^{-\gamma}$, cut-off $k_c \sim N^{1/2}$, and $f(k)=E_0 \log(k)$. In
the continuous degree approximation, this leads to
\begin{equation}
  \av{\tau} \simeq \int^{k_c} dk \; k^{1+\beta E_0-\gamma} \simeq
  k_c^{2+\beta E_0-\gamma}. 
\end{equation}
Since $H_{MFPT}$ is of order $N$ \cite{Redner:2001}, we obtain
\begin{equation}
  \av{t_{MFPT}} \simeq \left\{
    \begin{array}{lcl}
    N & \mathrm{if} & \beta E_0<\gamma-2 \\
    N^{2+(\beta E_0-\gamma)/2} & \mathrm{if} & \beta E_0>\gamma-2
  \end{array}
  \right. .
\label{eq:exponent_mod1}
\end{equation}
In the case of an exponential degree distribution,
%\begin{equation}
$$  
\av{\tau} \simeq \int^{k_c} dk  k e^{(\beta E_0-1)k/m} \simeq
  \frac{(\beta E_0-1)\frac{k_c}{m} -1}{\frac{(\beta E_0-1)^2}{m^2}} 
e^{(\beta E_0-1)\frac{k_c}{m}} \ , 
$$
%\end{equation}
\begin{figure}[t]
%\begin{center}
\includegraphics[width=0.42\textwidth]{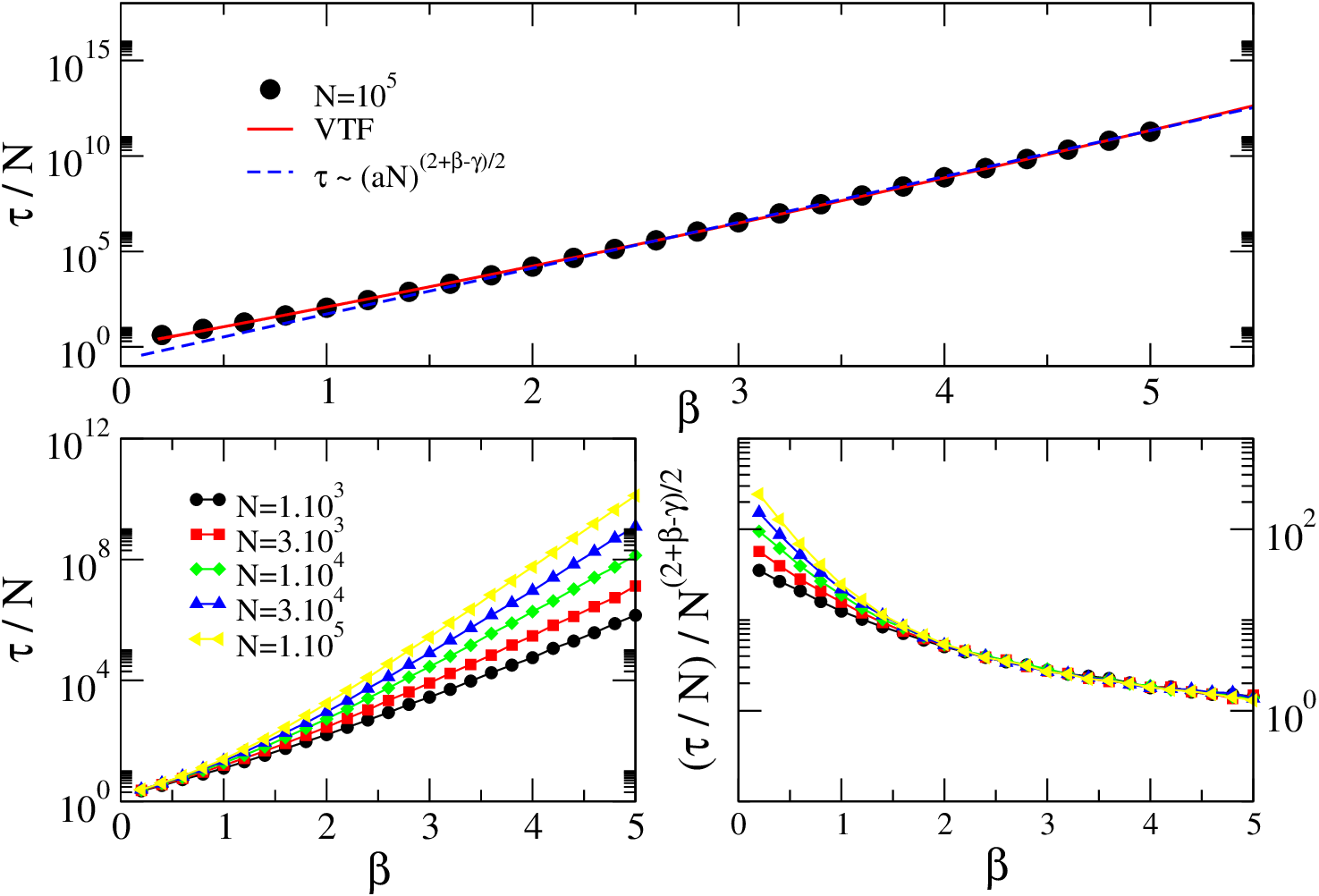}
%\end{center}
\caption{MFPT for scale-free uncorrelated random networks. Here
  $E_0=1$.  Top: $\gamma=2.2$, $N=10^5$; both Eq.(\ref{eq:exponent_mod1})
  and a VTF fit (with $T_0 \approx 0.023$) are shown. Bottom: $\gamma=3$ and
  various network sizes.  For $\beta < \beta_c=1$, $\tau\propto N$, while $\tau\propto N^{2+(\beta
    E_0-\gamma)/2}$ for $\beta > \beta_c$ .  }
\label{fig:MPFT}
\end{figure}
\noindent and, using $k_c \sim m \log N$, we obtain $\av{t_{MFPT}}
\simeq N$ for $\beta E_0 < 1$ and $\av{t_{MFPT}} \simeq N^{\beta E_0}$
for $\beta E_0 > 1$. Figure~\ref{fig:MPFT} shows the comparison of
numerical data with the prediction of
Eq. (\ref{eq:exponent_mod1}). The top panel also shows how,
interestingly, a Vogel-Tammann-Fulcher form $\exp(A/(T-T_0))$ can also
fit the data; however, the value of $T_0 \sim 0.023$ has here no clear
significance, while Eq. (\ref{eq:exponent_mod1}) provides a
straightforward interpretation of the data.

In summary, we have put forward a simple mathematical model for the
dynamics of glassy systems, seen as a random walk in a complex energy
landscape. This work puts previous studies on the topology of the
network of minima in a broader perspective and represents a first step
towards a systematic integration of tools and concepts developed in
complex network theory to the description of glassy dynamics in terms
of the exploration of a phase space seen as a network of minima. It
opens the way to studies on how network structures (such as community
structures or bottlenecks, large clustering, non-trivial correlations)
impact the dynamics. Other possible modifications of our model include
taking into account fluctuations of the energies within a degree class
(using for instance conditional energy distributions $P(E|k)$ instead
of a relation $E=f(k)$), and other transition rates $r(k \to k')$. A
preliminary analysis shows that, for Glauber rates, the same
phenomenology and the same necessary interplay between energy and
degree described here are obtained.  We also hope that this work will
stimulate further detailed investigations on the relation between
minima depth and connectivity.

\textit{Acknowledgments} A. Baronchelli
and R. P.-S.  acknowledge financial support from the Spanish MEC
(FEDER), under project No. FIS2007-66485-C02-01, as well as additional
support through ICREA Academia, funded by the Generalitat de
Catalunya.  A. Baronchelli aknowledges support of Spanish MCI through
the Juan de la Cierva programme.

\end{document}